# In A Society of Strangers, Kin Is Still Key: Identified Family Relations In Large-Scale Mobile Phone Data

Tamás Dávid-Barrett[a,b,c]
Sebastian Diaz[b,d]
Carlos Rodriguez-Sickert[a]
Isabel Behncke[a]
Anna Rotkirch[c]
János Kertész[e]
Loreto Bravo[d]

[a] Universidad del Desarrollo, Facultad de Gobierno, CICS, Av. Plaza 680, San Carlos de Apoquindo, Las Condes, Santiago de Chile, 7610658 Chile
[b] Trinity College, University of Oxford, OX1 3BH, Oxford, UK
[c] Population Research Institute, Väestöliitto, Kalevankatu 16, Helsinki 00101, Finland
[d] Data Science Institute, Universidad de Desarollo, Av. Plaza 680, Las Condes, 7610658, Santiago de Chile, Chile
[e] Central European University, Department of Network and Data Science, Nador u. 9, Budapest, H-1051, Hungary

**Abstract**

Mobile call networks have been widely used to investigate communication patterns and the network of interactions of humans at the societal scale. Yet, more detailed analysis is often hindered by having no information about the nature of the relationships, even if some metadata about the individuals are available. Using a unique, large mobile phone database with information about individual surnames in a population in which people inherit two surnames: one from their father, and one from their mother, we are able to differentiate among close kin relationship types. Here we focus on the difference between the most frequently called alters depending on whether they are family relationships or not. We find support in the data for two hypotheses: (1) phone calls between family members are more frequent and last longer than phone calls between non-kin, and (2) the phone call pattern between family members show a higher variation depending on the stage of life-course compared to non-family members. We give an interpretation of these findings within the framework of evolutionary anthropology: kinship matters even when demographic processes, such as low fertility, urbanisation and migration reduce the access to family members. Furthermore, our results provide tools for distinguishing between different kinds of kin relationships from mobile call data, when information about names are unavailable.

## Introduction

In all human cultures, people live in intensely social environments. The basis of every human society is multi-generational, multi-male-multi-female groups with strong pair bonds (Shultz et al., 2011). Throughout their lives, individuals build and maintain a particular set of social relations (Elder, 1994): to close kin, more distant kin, non-kin peers, mating partners, and their respective kin and peers (Kahn, 1980; Wrzus et al., 2013).

In high fertility, small-scale traditional societies which have characterised most of human evolutionary history, individuals tended to be surrounded by a network made up predominantly by relatives (Allen, 1989; Berté, 1988; Hames, 1987; Panter-brick, 1989), while low fertility and urbanised modern societies consist mostly of social networks dominated by friends and acquaintances (David-Barrett, 2019, 2020, 2022; David-Barrett and Dunbar, 2017; Hruschka, 2010). In both cases, frequent and meaningful social interactions play a key role in maintaining social bonds and enabling collaboration (Pollet et al., 2013; Roberts and Dunbar, 2011).

How we use our social environment through our life course has been studied by both biological sciences in evolutionary life history theory (Geary and Flinn, 2001; Geary et al., 2003; Hall, 2011; Hill and Kaplan, 1999; Rose and Rudolph, 2006), and social sciences in life course studies (Elder, 1994; Hutchison, 2015). Behavioural studies of close social bonds indicate that humans prefer to cooperate with kin rather than non-kin (Betzig and Turke, 1986; Burton-Chellew and Dunbar, 2011; Curry and Dunbar, 2013; Essock-Vitale and McGuire, 1985; Hughes, 1988; Johnson and Johnson, 1991; Morgan, 1979; Pollet et al., 2013; Sear and Mace, 2008; Shavit et al., 1994), similar to other social mammals (Archie et al., 2011; Creel and Creel, 2002; Krutzen et al., 2003) and in line with inclusive fitness theory (Hamilton, 1964a; Hamilton, 1964b). Between family generations, support is often altruistic and likely to flow from the older to the younger generations (Hughes, 1988). Family bonds are experienced as given before and beyond any conscious and deliberate individual act (Torche & Valenzuela, 2011). In contrast, compared to kin, relationships among just friends demand more reciprocal helping and higher frequency of contact in order to build and maintain relationship strength. Reciprocity and frequent contacts characteristic of friendships are likely to enhance trust and bonding, thus partly compensating for the lower shared genetic interest among non-kin (Rotkirch et al., 2014; Trivers, 1971), although friends are often low level kin with genetic relatedness higher than the population average (Christakis and Fowler, 2014), but lower than kin assignment via linguistic kin terms (Dunbar, 1997).

Recent access to large communication data, especially to mobile phone Call Detailed Records (CDR-s) provides unprecedented opportunities to study social relationships at a societal scale (Onnela et al., 2007, Blondel, 2015). An important limitation of the usage of such data is that usually no information about the type of the links are available, even if metadata about gender, age, location of the individuals are known.

Recent research has attempted to identify kin and peer relationship types from anonymous mobile communication data, based on a combination of phone call patterns and basic socio-demographic information (age and gender) of the callers, drawn from a contemporary European population (David-Barrett et al., 2016b; Palchykov et al., 2012). This new methodology exploited the fact that the average mobile phone caller has up to six distinct peaks in the histogram of call frequency (David-Barrett et al., 2016b), as a function of the alters' gender and age relative to the caller. The positions of these peaks are conspicuous (David-Barrett et al., 2016b): two correspond with alters who are approximately one generation older. In these, the male most frequently called alter is a few years older on average than the female. These were independent of the ego's sex. The assumption of that paper was that these two alters may correspond to the mother and the father of the ego.

In the same-generation peaks, both the sex of the ego and the sex of the alter mattered. If the sex is different, then on the average the male was a few years older than the female. If the sex was the same, there was no age difference. The paper suggested the assumption that in the former case the relationships are predominantly of romantic nature, while in the latter case these correspond mostly to best same-sex friend relationships.

The one-generation-younger peaks showed no sex difference in age. The assumption was that these were most likely the children of the ego.

The life course pattern of other phone call characteristics, for instance direction of call initiation and length of phone calls were consistent with this hypothesis, providing further support to it (David-Barrett et al., 2016b).

This methodology allowed a refined differentiation among close kin networks. However, these assignments of relationship types were hypothesised. For instance, while it is likely that the average person's most frequently called female alter who is one generation older is the person's mother, it is impossible to tell for sure.

Here, we are able to move beyond this methodological limitation using a large phone database uniquely tagged with information about individual surnames. These are hashed, satisfying GDPR requirements, so that we can assure whether two names are identical, but we do not have access to the real names. Using surnames adds significantly more information, since for instance two full brothers are unlikely to have different surnames while two male friends are unlikely to have the same paternal and maternal surnames. This is especially so, since the data is from Chile, as in many Spanish-speaking countries, both the patrilineal paternal and matrilineal paternal names are part of the family name. Individuals, thus, have two family names: the first is the first family name of the father, and the second is the first family name of the mother. Phone companies record both family names of an overwhelming majority of their clients, enabling a refined detection of the nature of dyadic kin relationships in the communication database.

Making use of the information on surnames contained in the Chilean mobile phone data, we are able to distinguish between real relatives and "quasi" relatives. For instance, a one-generation-older female alter will be classified as the ego's mother only if the second name of the ego coincides with the first name of the alter. If this is not the case, we will refer to this alter as a "quasi mother". Our focus here is the set of cross-generational close family relationships: mother, father, daughter, son.

On the basis of the distinction between relatives and quasi relatives for the cross-generational relationships, we study the differences in the way subjects interact with kin and frequently called non-related individuals who, given their age and gender, could perform a similar role in the life of the individual. Thus, we compare, for instance, within the category of most frequently called one-generation-older female, depending on whether they are the mother of the ego or the most frequently called mother-aged female who is not the mother. We think of this as a comparison of the most important mother figure depending on whether she happens to be the real mother of the ego or a quasi-mother.
We study call patterns through four variables frequently used in previous studies: (i) frequency of calls for a specific dyad, (ii) fraction of total phone call time within a specific dyad compared to ego's total call time, (iii) out-call fraction: the proportion of outgoing calls, relative to the sum of incoming and outgoing calls within a specific dyad which represents the balance in the relationship, and (iv) the average call length (David-Barrett et al., 2016b).

Our first hypothesis concerns the intensity of the relationship -frequency and call length- with a close biological kin versus with a non-kin, even if they occupy a similar role in the social network of the ego.

H1. If the most frequently called persons within a kin-category (e.g., mother, father, daughter, son) are genetic relationships, then we expect that the frequency, call length and fraction of time are higher compared to the quasi relationships. For instance, if we are comparing the egos' call patterns with the most frequently called one-generation-older females, we expect that if these women are the biological mother of the egos, the frequency is higher compared to the unrelated quasi-mothers.

Our second hypothesis is based on the first. The higher intensity between biologically related versus non-related social contacts manifests itself especially in situations where one relies on the other for help or cooperation. These situations are not distributed evenly through life. Consider for instance the difference between the relationship of a daughter with her real mother and a quasi-mother when the daughter has her own offspring. Given the evolutionary explanation of females survival past the end of reproduction is in general associated with grandmothering (Cant & Croft, 2019), one would expect a higher relative peak during this period for the real mother.

H2. We expect that the life-course dependent variation of all the phone call patterns (frequency, fraction of time, call length and also out-call fraction) is higher for real relatives than apparent relatives. For instance, in the case of the mother figure, we expect that the life-course dependent variation of the direction of phone call initiation (who calls whom), or the length of the phone call, is higher between the ego and the mother figure, if the latter is the actual rather than the quasi mother of the ego.

Note that we are not able to carry out a longitudinal study due to the limited time span of the data. When we refer to life-course dependent variations, we mean that the characteristics in different age groups show systematic variations, and thus we assume that the cohort effects are significantly smaller than the effect of life-course.

## Data and Methods

The data is aggregated anonymized Call Detail Records (CDRs) from a Chilean Mobile Call Company accounting for 40% of the market share for the respective period. CDRs are generated automatically by the telephone company every time that a call is made or received by a person inside the network. This data is collected automatically for billing purposes.

Each record stores, among other things, the origin and destination number and antenna, a time stamp (day, hour, minute, second) and the duration of the call in seconds. The data used in this study was anonymized using several techniques (see the supplementary material for details).

The data was collected for the 12 consecutive months of the year 2015, totaling 3,994,595,128 calls. Using all this data we created a graph where each node corresponds to a phone number, and there is an edge between two phone numbers if there is at least one phone call between them. The final graph created consisted in 8,907,140 vertices and 112,744,511 edges.

Ethical permission for this project was granted by the Ethics Committee of Universidad del Desarrollo under the code CEII09-2019.

Given an ego and an alter, we study call patterns through four variables: (i) **Frequency**: frequency of calls between them, (ii) **FracOfTime**: fraction of total phone call time within the specific dyad compared to ego's total call time, (iii) **OutCallFrac**: the proportion of outgoing calls, relative to the sum of incoming and outgoing calls within the specific dyad, and (iv) **CallLength**: the average time per call.

For identifying the close kin ego network, we here developed a variant of the David-Barrett et al. (David-Barrett et al., 2016b) methodology. The original methodology allowed the identification of ego's most important male and female contacts in one generation older groups, same-generation groups, and one generation younger groups, as outlined above in the Introduction. It hypothesised that these generations were likely to represent the mother, father, romantic partner, best friend, daughter, and son of the ego. Considering that we can only identify the age and gender of people within the same phone company, this identification naturally introduces considerable "false positive" type errors. In this paper we introduce filters based on further available information about the users, which substantially increases the accuracy of the estimation of the types of social and family relationships.

The metadata collected by the company includes besides caller and alter age, gender already used in (David-Barrett et al., 2016b) also anonymised last names, i.e., patrilineal paternity *and* the matrilineal paternity names, which enable us to construct such filters.

The categories were created based on the definitions of David-Barrett et al. (David-Barrett et al., 2016b) methodology, which detects candidate relationship categories by applying three filters: 1) a demographic filter (generation and gender) and 2) a call frequency filter. For each category, we add an additional 3) "surname" filter that uses patrilineal and matrilineal family names that separates the candidate mothers.

Thus, we used three filters to define the categories. First, we applied a demographic filter, which defined the position of each alter compared to the ego in terms of relative generations. This yielded sets of alters for each ego, who were an older generation, same generation, or younger generation. In each generation set, we separated the female and the male alters. Thus, this filter partitioned egos' alters into six subsets based on relative age and sex.

Second, we used a call frequency filter for each subset of alters for each ego. This allowed us to identify the one alter in each subset (provided that the subset exists) with whom the ego conducted calls the most frequently.

Third, we used a surname filter where we compared the family names of the ego. A child takes as first last name the first last name of his/her father, and as second last name the first last name of his/her mother. If this last name rule applies, then we denote the selected alter as kin, for instance the father, and if they did not match then we denoted it as quasi-kin, for instance, the quasi-father.

Note that even with the use of the last name filter, it is still possible that some of the relationship types are misidentified (see Table 1.) For instance, an ego may have the most frequent phone call with a one-generation-older woman with whom ego shares the last name both in the case of the mother, or the mother's sister, i.e., the ego's maternal aunt. Similarly, our assumptions leave room for false negatives, in cases where the kin relationship is not the most frequently called. The generation age grouping, frequent surnames and the fact that we can detect at most two kids (one daughter and one son), can also be a source of both false positives and false negatives.

Thus, in the case of calls of mothers we first applied the demographic and call frequency filters in which we took the most frequently called female alter among all alters with ages of 15-40 years older than the ego. We partitioned this group in two:

a) Mother: The subgroup of alters that shared their first last names with the second last names of the ego. This filter will thus with great certainty pick up actual mothers (and some maternal aunts, see Table 1).

b) Quasi-mother: The subgroup that did not share their first last names with the second last names of the ego. We assume that these are mother figures who are not the real mother. (Note that a stepmother or a paternal aunt would be categorised here, as well as non-related mother figures.)

For each category, we created the same two subcategories with the corresponding filters.

The 'candidate father' is the most frequently called male alter among those who are 17-42 years older than the ego. If the first last name of the ego matches the first last name of the alter, then we consider it to be the 'father'. (Note: this could include some paternal uncles). If this is not the case, then we label it as quasi-father.

The 'candidate daughter' is defined as the most frequently called female alter one generation younger than the ego. This is 17-42 years older for male egos and 15-40 for the female egos. If in addition the first last name of the ego matches first last name of the alter for the male egos (fathers) and first last name of the ego matches the second last name of the alter for female egos (mothers) we consider her to be the 'daughter'. If this is not the case, we consider her to be the quasi-daughter. Female siblings of the candidate daughter and of the quasi-daughter are not considered in the analysis.

The 'candidate son' is the most frequently called male alter one generation younger than the ego is called 'son'. This is 17-42 years older for male egos and 15-40 for the female egos. If in addition the first last name of the ego matches first last name of the alter for the male egos (fathers) and first last name of the ego matches the second last name of the alter for female egos (mothers) we consider him to be 'son' If this is not the case, we consider him to be quasi-son. Female siblings of the candidate son and of the quasi-son are neither considered in the analysis.

**Table 1. Relationship definitions**

|  | **Demographic and call frequency filter** | **Surname filter** | **Possible misidentification** |
|---|---|---|---|
| Mother | Most frequent female contact 15-40 years older | Second last name of the ego is the same as the first last name of the alter. | Maternal aunt |
| Father | Most frequent male contact 17-42 years older | First last name of the ego is the same as the first last name of the alter. | Paternal uncle |
| Daughter | Most frequent female contact 15-40 (17-42) years younger for female (male) ego | First last name of the ego is the same as the first(male ego) or second last name(female ego) of the alter. | Niece (sister's daughter) |
| Son | Most frequent male contact 15-40 (17-42) years younger for female (male) ego | First last name of the ego is the same as the first(male ego) or second last name(female ego) of the alter. | Nephew (brother's son) |

Consider the example in Fig. 1. We have 4 females (in red) and 2 males (in blue) where each node is labelled with the two last names and age. The edges are labelled by the total number of phone calls.

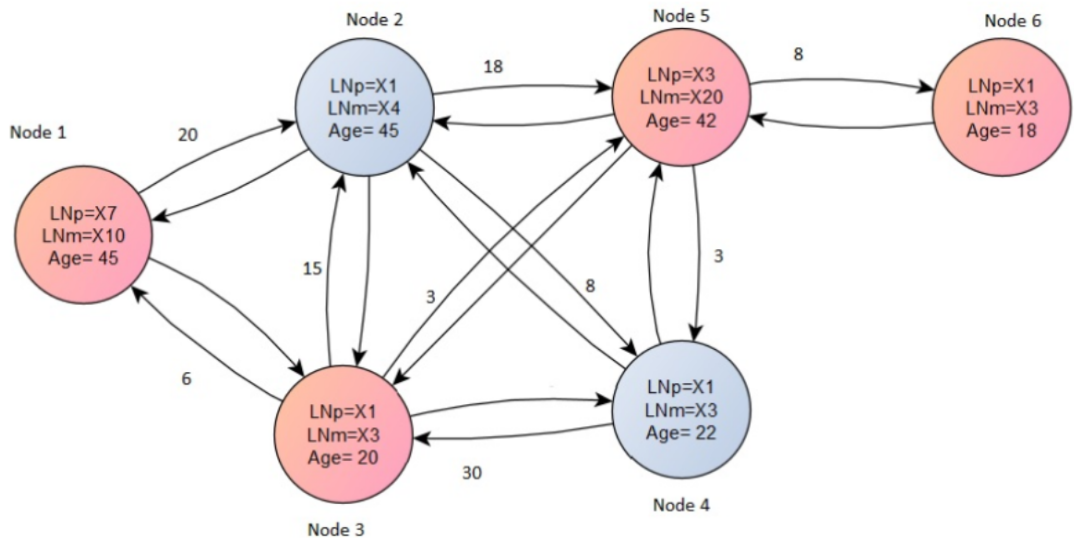

Fig. 1. Example of a communication network among kin where LNp denotes paternal last name, and LNm maternal last name. Ages are indicated. Red circles: females; blue circles: males. The numbers near the arrows are the number of calls within the dyad.

In this example, Node 1 is a quasi-mother of Node 3 (and conversely, node 3 is the quasi-daughter of Node 1). Node 2 is the father of Node 4 and Node 4 is the son of Node 2. Finally, Node 5's daughter is Node 6 and Node 6's mother is Node 5.

## Results

Our aim is to investigate differences in call patterns between a specific social alter and other alters of the same age group and gender.

First, we compared phone calls between egos and their mothers and egos and their quasi-mothers (see Fig.3). Our results show that there is a marked difference between phone calls with those that are identified as mothers and quasi-mothers In particular, for both female and male egos (i) the frequency of phone calls to mothers compared to quasi-mothers is higher at all ego ages; and (ii) apart from early 20s and 60+ egos, for all other ego ages the length of the phone call between the ego and the mother is longer than calls to quasi mothers.

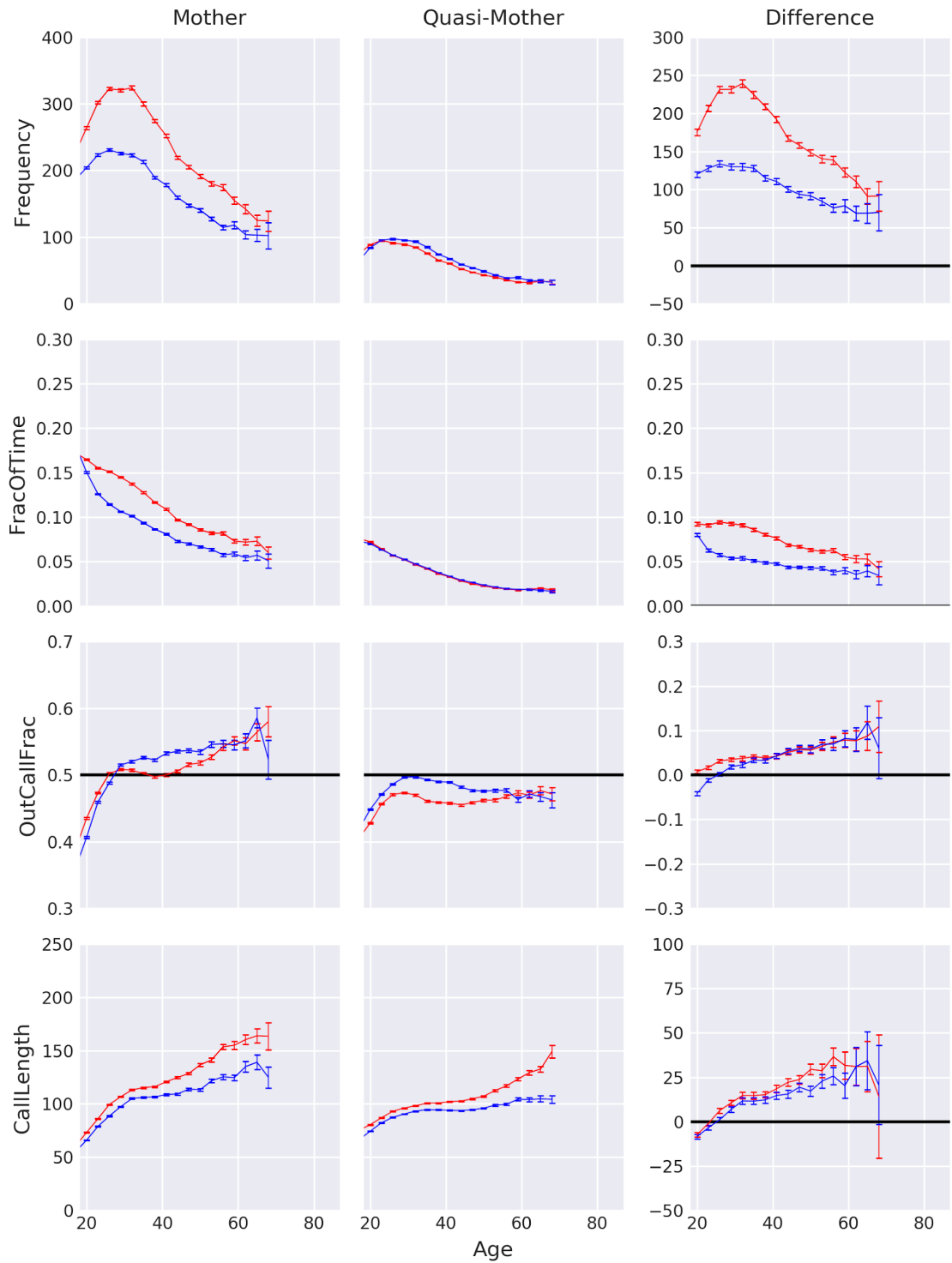

Fig. 3. Call patterns (Frequency, FracOfTime, OutCallFrac and CallLength) between egos and their mothers (first column) and quasi-mothers (second column). The third column represents the difference between the mother and quasi-mother. The red and blue lines correspond to female and male egos respectively

Furthermore, there is a particular age-dependent pattern when comparing out-call fraction and call length. The majority of phone calls to mothers during the ego's early 20s are initiated by their mother (up to 60%). In this period the phone calls between mother and adult child are also short. This pattern changes significantly by the second half of the ego's 20s, when the ego becomes much more likely to initiate phone calls and when the average length of the calls doubles.

There is a significant sex difference: female egos are more likely to have frequent phone conversations and longer phone calls with their mother than male egos, independent of age. In the case of the quasi-mother alters, a similar pattern can be seen for both female and male egos. However we observe a difference in the out-call fraction. Whereas, for 20 year old egos, both mothers and quasi mothers more frequently initiate their calls, this relation is reversed only for the mothers once the ego reaches late 20's-early 30's. For quasi-mothers this shift in call initiation does not occur.

These results are consistent with the earlier suggestion by David-Barrett et al. (David-Barrett et al., 2016b) that the shift in phone calls is most likely to be explained by increasing reliance on grandmaternal care. Mid-20s is indeed the mean mother's age at first birth in Chile. This would be consistent with the fact that there is a strong sex difference in call frequency to mothers at this age, not seen in alters of the same socio-demographic group who were quasi-mothers, that the initiation by the ego is most likely to take place at this period, also with a strong sex difference, and that the length of phone calls increase to a larger extent to mothers than to quasi-mother in this period.

An additional finding is that, in the case of non-kin the majority outcalls are made by the older ego, independent of the age. This unbalanced reciprocity/direction might be explained by several factors, including additional spare time for women. Women retire earlier and have a labor force participation of 50% vs 75% of men. This inequality is higher in the case of males for kin. But the opposite is true for non-kin.

In the case of fathers compared to quasi-father (Fig. 4) callers in the same generation, the overall pattern is similar to mothers vs. quasi-mothers. Both the call frequency is higher, and the call lengths are longer with fathers than with quasi-fathers.

However, there are some notable differences as well. First, the difference between frequently called fathers and frequently called quasi-fathers is smaller than in the case of mothers and quasi-mothers. Second, the strong gender difference observed in the case of mothers vs. quasi-mothers is muted in the case of fathers vs. quasi-fathers.

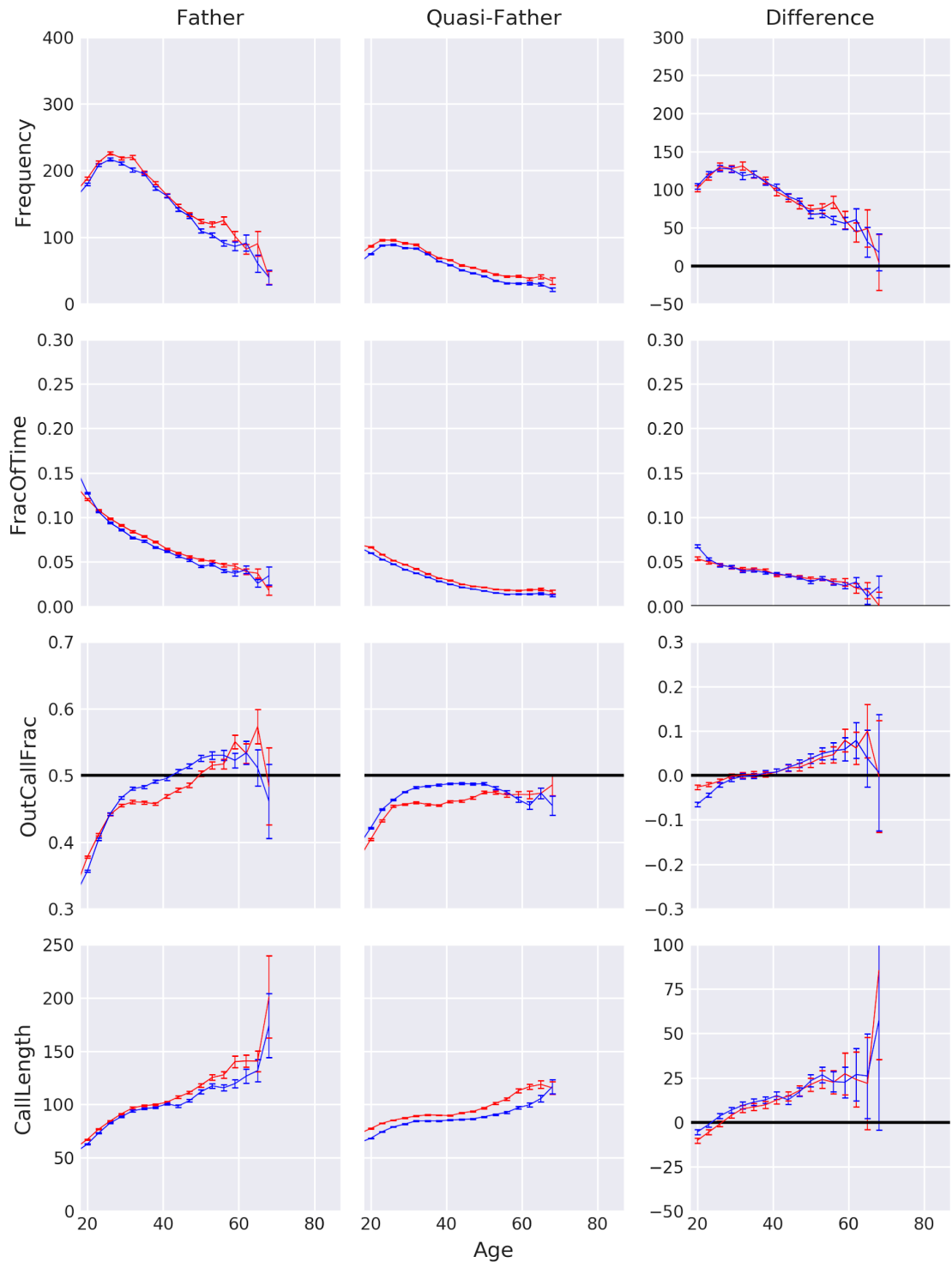

Fig. 4. Call patterns (Frequency, FracOfTime, OutCallFrac and CallLength) between egos and their fathers (first column) and quasi-fathers (second column). The third column represents the difference between the father and quasi-father. The red and blue lines correspond to female and male egos respectively

The pattern that we saw with mothers and fathers as compared to quasi-parents is further repeated in calls with those that were identified as daughters vs. quasi-daughters (Fig. 4). In particular, the frequency of calls with daughters is higher than to quasi-daughter most frequent callers, and last longer. The sex difference in call pattern is also confirmed:

mothers are more likely to have frequent and long phone calls with their daughter than are fathers.

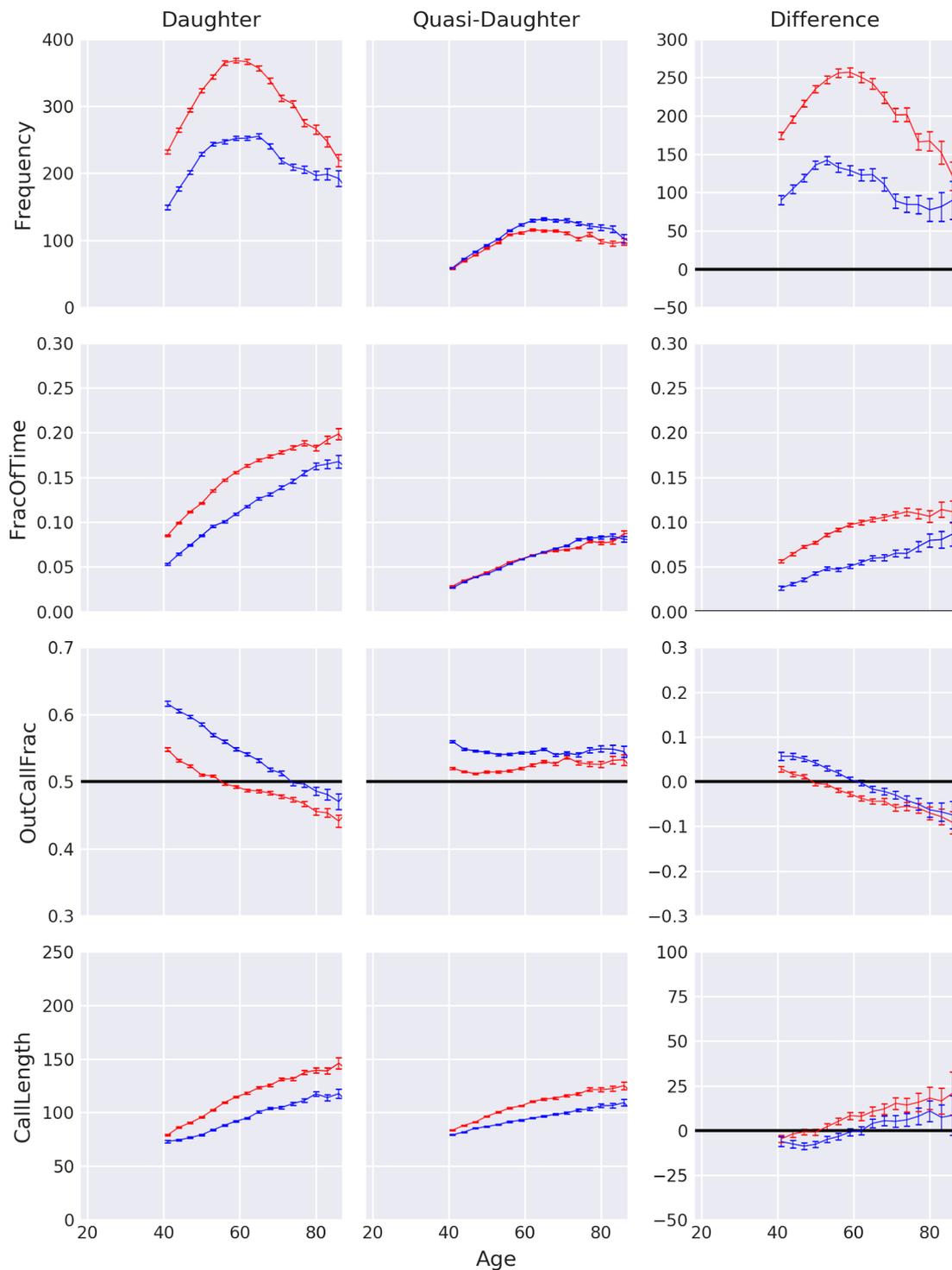

Fig. 5. Call patterns (Frequency, FracOfTime, OutCallFrac and CallLength) between egos and their daughter (first column) and quasi-daughter (second column). The third column represents the difference between the daughter and quasi-daughter. The red and blue lines correspond to female and male egos respectively.

There is a similar pattern with sons as with daughters above, however, the difference between sons vs. quasi-sons is smaller (Fig. 5).

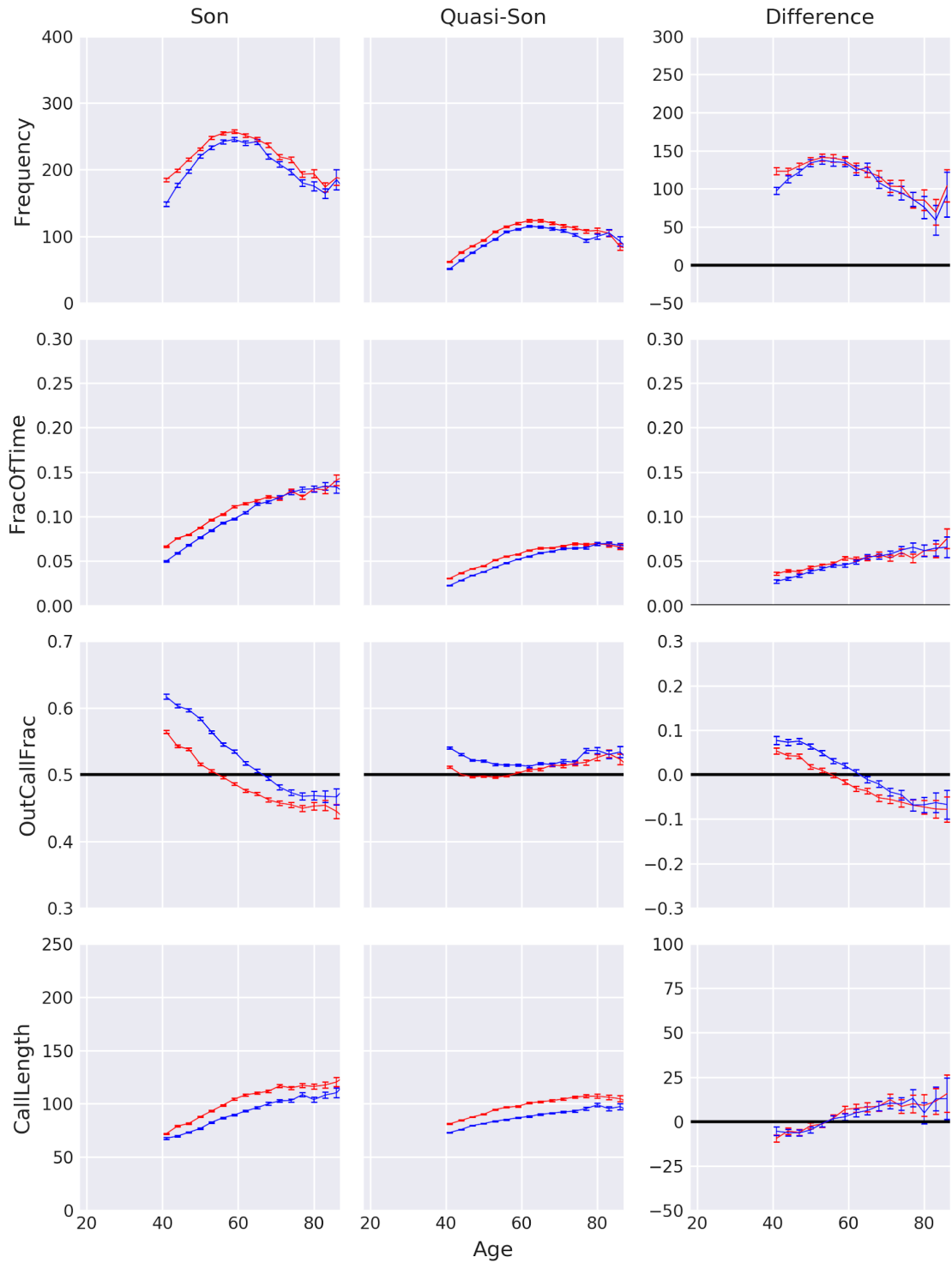

Fig. 5. Call patterns (Frequency, FracOfTime, OutCallFrac and CallLength) between egos and their son (first column) and quasi-son (second column). The third column represents the difference between the son and quasi-son . The red and blue lines correspond to female and male egos respectively.
.

We also measured the life-course dependent variation of the phone call pattern in the case of frequency, fraction of time, out call fraction, length of phone call (Table S1, is supplementary material). For every variable we observe higher variation for every category of kin compared to quasi-kin. (eg. male father, vs male quasi father variation). This gives support to our second hypothesis. The higher intensity between biologically related versus non-related social contacts manifests itself especially in situations where one relies on the other for help or cooperation, e.g daughters having their own offspring (see first row, in fig. 3) where the peak in frequency and difference with non-kin to mothers occur during that period 24-40 years. Our results are consistent with the fact that these situations take place in a specific period of the life cycle.

## Discussion

In this paper, we have presented results from a mobile phone communication network, using reliable and high-quality data from contemporary Chile, that compared specific biological kin relationships with non-kin links that appear in a similar position in the ego networks. Our results suggest that the previous attempts to identify family ties and the related behaviour went in a proper direction (David-Barrett et al., 2016b), however, using the opportunity given by the unique Chilean data, it is possible to test the hypotheses even closer. We have confirmed that the previously found effects are present and even more pronounced in genetically related ties compared to non-related ties. Thus, we have shown that whether a real family relationship is underlying the mobile phone interaction between ego and alter is important in the key characteristics of the interaction.

When comparing the cases when the most frequent calling partner is a relative, versus is a non-relative, we find that the call frequency is higher, the share among all the phone calls is higher, and the length of the calls is higher. This confirmed our first hypothesis. We have also shown that the life course dependent variation of the relationship in the call initiation direction is higher when the most frequent caller is relative compared to a non-relative. This confirmed our second hypothesis.

Furthermore, we have found that the difference between mothers and quasi-mothers is muted between fathers and quasi-fathers. This finding is in line with the fact that women are more likely to maintain the cross generational bond within a family (David-Barrett et al., 2016b). They may also reflect the effects of divorce and remarriage.

That genetic relatedness matters is far from being a new finding (Bowles and Posel, 2005; Danielsbacka et al., 2015; Tanskanen et al., 2016; Wolf et al., 2011), violated only in a very few relationships of any individual (David-Barrett et al., 2015). The importance of life-course phases in social behaviour has also been established in the context of social networks (David-Barrett et al., 2016a; Hooper et al., 2015; Kalmijn, 2003). However, to our knowledge, this is the first study that has differentiated between social dyads of maximum-level importance depending on the presence or absence of kin relatedness. This result contributes to our understanding of how social network edges vary along the underlying type of relationship as well as the phase of the life-course that the interacting parties are in.

# REFERENCES

Albert, R., Barabasi, A. L., 2002. Statistical mechanics of complex networks. Reviews of Modern Physics 74, 47-97.

Allen, N. J., 1989. The Evolution of Kinship Terminologies. Lingua 77, 173-185.

Archie, E. A., Moss, C. J., Alberts, S. C., 2011. Friends and relations: kinship and the nature of female elephant social relationships. In: Moss, C. J., Croze, H., Eds.), Amboseli Elephants: A long-term perspective on a long-lived mammal University of Chicago Press, Chicago.

Barabási, A.-L., 2002. Linked : the new science of networks. Perseus Publishing, Cambridge, Mass.

Barabási, A.-L., Pósfai, M., 2016. Network science. Cambridge University Press, Cambridge.

Berté, N. A., 1988. K'ekchi' horticultural labor exchange: Productive and reproductive implications. In: Betzig, L., et al., Eds.), Human reproductive behaviour Cambridge University Press., Cambridge, pp. 83–96.

Betzig, L., Turke, P., 1986. Food sharing on Ifaluk. Current Anthropology 27, 397–400.

Blondel, V. D., Decuyper, A., Krings, G., 2015. A survey of results on mobile phone datasets analysis, EPJ Data Science, 4, Article number: 10.

Bowles, S., Posel, D., 2005. Genetic relatedness predicts South African migrant workers' remittances to their families. Nature 434, 380-383, doi:10.1038/nature03420.

Burton-Chellew, M. N., Dunbar, R. I. M., 2011. Are Affines Treated as Biological Kin? A Test of Hughes's Hypothesis. Current Anthropology 52, 741-746, doi:10.1086/661288.

Cant, M. A., & Croft, D. P. (2019). Life-History Evolution: Grandmothering in Space and Time. Current Biology, 29(6), R215-R218.

Christakis, N. A., Fowler, J. H., 2014. Friendship and natural selection. Proceedings of the National Academy of Sciences of the United States of America 111, 10796-10801, doi:10.1073/pnas.1400825111.

Creel, S., Creel, N. M., 2002. The African wild dog : behavior, ecology, and conservation. Princeton University Press, Princeton, N.J.; Oxford.

Curry, O., Dunbar, R. I., 2013. Do birds of a feather flock together? The relationship between similarity and altruism in social networks. Hum Nat 24, 336-47, doi:10.1007/s12110-013-9174-z.

Danielsbacka, M., Tanskanen, A. O., Rotkirch, A., 2015. Impact of Genetic Relatedness and Emotional Closeness on Intergenerational Relations. Journal of Marriage and Family.

David-Barrett, T., 2019. Network Effects of Demographic Transition. Scientific Reports 9, doi:10.1038/s41598-019-39025-4.

David-Barrett, T., 2020. Herding Friends in Similarity-Based Architecture of Social Networks. Scientific Reports 10, 4859. doi: 10.1038/s41598-020-61330-6

David-Barrett, T., 2022. David-Barrett, T. (2022) Kinship Is a Network Tracking Social Technology, Not an Evolutionary Phenomenon, arXiv 2203.02964. doi:10.48550/arXiv.2203.02964.

David-Barrett, T., Dunbar, R. I. M., 2017. Fertility, kinship, and the evolution of mass ideologies. J Theor Biol 417, 20-27, doi:10.1016/j.jtbi.2017.01.015

David-Barrett, T., Behncke Izquierdo, I., Carney, J., Nowak, K., Launay, J., Rotkirch, A., 2016a. Life Course Similarities on Social Network Sites. Advances in Life Course Research

David-Barrett, T., Kertesz, J., Rotkirch, A., Ghosh, A., Bhattacharya, K., Monsivais, D., Kaski, K., 2016b. Communication with Family and Friends across the Life Course. Plos One 11, doi:ARTN e016510.1371/journal.pone.0165687.

David-Barrett, T., Rotkirch, A., Carney, J., Behncke Izquierdo, I., Krems, J. A., Townley, D., McDaniell, E., Byrne-Smith, A., Dunbar, R. I., 2015. Women favour dyadic relationships, but men prefer clubs: cross-cultural evidence from social networking. PLoS One 10, e0118329, doi:10.1371/journal.pone.0118329.

Dodds, P. S., Muhamad, R., Watts, D. J., 2003. An experimental study of search in global social networks. Science 301, 827-9, doi:10.1126/science.1081058.

Dunbar, R. I. M., 1997. On the evolution of language and kinship. In: Steele, J., Shennan, S., Eds.), The Archaeology of Human Ancestry: Power, Sex and Tradition. Routledge, London, pp. 380-396.

Elder, G. H., 1994. Time, Human Agency, and Social-Change - Perspectives on the Life-Course. Social Psychology Quarterly 57, 4-15.

Erdos, P., Renyi, A., 1960. On the Evolution of Random Graphs. Bulletin of the International Statistical Institute 38, 343-347.

Essock-Vitale, S. M., McGuire, M. T., 1985. Women's lives viewed from an evolutionary perspective: II. Patterns of helping. Ethology and Sociobiology 6, 155-173.

Euler, L., 1736. Solutio problematis ad geometriam situs pertinentis. Comment. Acad. Sci. U. Petrop 8, 128-140.

Freeman, L. C., 2004. The development of social network analysis : a study in the sociology of science. Empirical Press, Vancouver, BC.

Geary, D. C., Flinn, M. V., 2001. Evolution of Human Parental Behavior and the Human Family. Parenting-Science and Practice 1, 5-61, doi:Pii 785828753

Doi 10.1207/S15327922par011&2_2.

Geary, D. C., Byrd-Craven, J., Hoard, M. K., Vigil, J., Numtee, C., 2003. Evolution and development of boys' social behavior. Developmental Review 23, 444-470, doi:Doi 10.1016/J.Dr.2003.08.001.

Hall, J. A., 2011. Sex differences in friendship expectations: A meta-analysis. Journal of Social and Personal Relationships 28, 723-747, doi:Doi 10.1177/0265407510386192.

Hames, R., 1987. Garden labor exchange among the Ye'Kwana. Ethology and Sociobiology 8, 259–284.

Hamilton, W. D., 1964a. Genetical Evolution of Social Behaviour I. Journal of Theoretical Biology 7, 1-&, doi:Doi 10.1016/0022-5193(64)90038-4.

Hamilton, W. D., 1964b. Genetical Evolution of Social Behaviour 2. Journal of Theoretical Biology 7, 17-&, doi:Doi 10.1016/0022-5193(64)90039-6.

Hill, K., Kaplan, H., 1999. Life history traits in humans: Theory and empirical studies. Annual Review of Anthropology 28, 397-430, doi:DOI 10.1146/annurev.anthro.28.1.397.

Hooper, P. L., Gurven, M., Winking, J., Kaplan, H. S., 2015. Inclusive fitness and differential productivity across the life course determine intergenerational transfers in a small-scale human society. Proc Biol Sci 282, doi:10.1098/rspb.2014.2808.

Hruschka, D. J., 2010. Friendship : development, ecology, and evolution of a relationship. University of California Press, Berkeley, Calif. ; London.

Hughes, A. L., 1988. Evolution and human kinship. Oxford University Press, New York ; Oxford.

Hutchison, E. D., 2015. Dimensions of human behavior : the changing life course. SAGE,, Los Angeles, pp. 1 online resource (xxvi, 542 pages).

Johnson, S. B., Johnson, R. C., 1991. Support and Conflict of Kinsmen in Norse Earldoms, Icelandic Families, and the English Royalty. Ethology and Sociobiology 12, 211-220, doi:Doi 10.1016/0162-3095(91)90004-A.

Kahn, R. L. A., T. C. , 1980. Convoys over the life course: attachment, roles and social support., in In: Baltes, P., Brim, O., Eds.), Life-span Development and Behavior. Academic Press, New York, pp. 253–286.

Kalmijn, M., 2003. Shared friendship networks and the life course: an analysis of survey data on married and cohabiting couples. Social Networks 25, 231-249, doi:10.1016/S0378-8733(03)00010-8.

Konig, D., 1936. Theorie der endlichen und unendlichen Graphen. Akademische Verlagsgesellschaft, Leipzig.

Krutzen, M., Sherwin, W. B., Connor, R. C., Barre, L. M., Van de Casteele, T., Mann, J., Brooks, R., 2003. Contrasting relatedness patterns in bottlenose dolphins (Tursiops sp.) with different alliance strategies. Proceedings of the Royal Society B-Biological Sciences 270, 497-502, doi:Doi 10.1098/Rspb.2002.2229.

Morgan, C. J., 1979. Eskimo hunting groups, social kinship and the possibility of kin selection in humans. Ethology and Sociobiology 1, 83-86.

Onnela, J.-P., Saramäki, J. Hyvönen, J., Szabó, G., Lazer, D., Kaski, K., Kertész, J., Barabási, A.-L., 2007. Structure and tie strengths in mobile communication networks, PNAS 104 (18) 7332-7336

Palchykov, V., Kaski, K., Kertesz, J., Barabasi, A. L., Dunbar, R. I., 2012. Sex differences in intimate relationships. Sci Rep 2, 370, doi:10.1038/srep00370.

Panter-brick, C., 1989. Motherhood and subsistence work: the Tamang of rural Nepal. Hum Ecol 17, 205-28.

Pollet, T. V., Roberts, S. G., Dunbar, R. I., 2013. Going that extra mile: individuals travel further to maintain face-to-face contact with highly related kin than with less related kin. PLoS One 8, e53929, doi:10.1371/journal.pone.0053929.

Roberts, S. G. B., Dunbar, R. I. M., 2011. The costs of family and friends: an 18-month longitudinal study of relationship maintenance and decay. Evolution and Human Behavior 32, 186-197, doi:10.1016/j.evolhumbehav.2010.08.005.

Rose, A. J., Rudolph, K. D., 2006. A review of sex differences in peer relationship processes: Potential trade-offs for the emotional and behavioral development of girls and boys. Psychological Bulletin 132, 98-131, doi:10.1037/0033-2909.132.1.98.

Rotkirch, A., Lyons, M., David-Barrett, T., Jokela, M., 2014. Gratitude for Help among Adult Friends and Siblings. Evolutionary Psychology 12, 673-686.

Ryan, B., Gross, N., 1943. The diffusion of hybrid seed corn in two Iowa communities. Rural Sociology 8, 15-24.

Scott, J., Carrington, P. J., 2011. The SAGE handbook of social network analysis. SAGE,, London ; Thousand Oaks, Calif., pp. 1 online resource (xvi, 622 pages).

Sear, R., Mace, R., 2008. Who keeps children alive? A review of the effects of kin on child survival. Evolution and Human Behavior 29, 1-18, doi:10.1016/j.evolhumbehav.2007.10.001.

Shavit, Y., Fischer, C. S., Koresh, Y., 1994. Kin and Nonkin under Collective Threat - Israeli Networks during the Gulf-War. Social Forces 72, 1197-1215, doi:Doi 10.2307/2580298.

Shultz, S., Opie, C., Atkinson, Q. D., 2011. Stepwise evolution of stable sociality in primates. Nature 479, 219-22, doi:10.1038/nature10601.

Tanskanen, A. O., Danielsbacka, M., Jokela, M., David-Barrett, T., Rotkirch, A., 2016. Diluted Competition? Conflicts between Full- and Half-Siblings in Two Adult Generations. Frontiers in Sociology 1, doi:10.3389/fsoc.2016.00006.

Torche, F., & Valenzuela, E. (2011). Trust and reciprocity: A theoretical distinction of the sources of social capital. European Journal of Social Theory, 14(2), 181-198.

Trivers, R. L., 1971. Evolution of Reciprocal Altruism. Quarterly Review of Biology 46, 35-&, doi:Doi 10.1086/406755.

Wolf, J. B., Traulsen, A., James, R., 2011. Exploring the Link between Genetic Relatedness r and Social Contact Structure k in Animal Social Networks. Am Nat 177, 135-42, doi:10.1086/657442.

Wrzus, C., Hanel, M., Wagner, J., Neyer, F. J., 2013. Social Network Changes and Life Events Across the Life Span: A Meta-Analysis. Psychological Bulletin 139, 53-80, doi:10.1037/a0028601.

# Supplementary Material
## Comparison with Previous Methodology

Applying the surname filters to the data only partly confirmed the validity of identifying the types of relationships between people based on age, sex, and frequency of phone calls introduced by David-Barrett et al. (David-Barrett et al., 2016b).

Fig 2. shows the percentage of kin identification, based on age, sex, and frequency of phone calls introduced by David-Barrett et al. (David-Barrett et al., 2016b), which is near 5-35%. This was expected, and a logical implication of working with censored data, where market share of the company is only 30%.

Children: a minority of the children identified from the frequency-age criterion were confirmed as the children of the egos based on surname identification. For female egos (i.e., assumed mothers of a child) 5-35% of assumed children were identified as filtered sons or daughters. For male egos (i.e., assumed fathers) the ratio is lower, especially for younger children, ranging between 5 and 25%.

Parents: the majority of alters that were identified using the frequency-age criteria as parents were not confirmed as the parents based on surname data. Mothers were confirmed in 3-30% of cases, while fathers were confirmed 1-20% of the time.

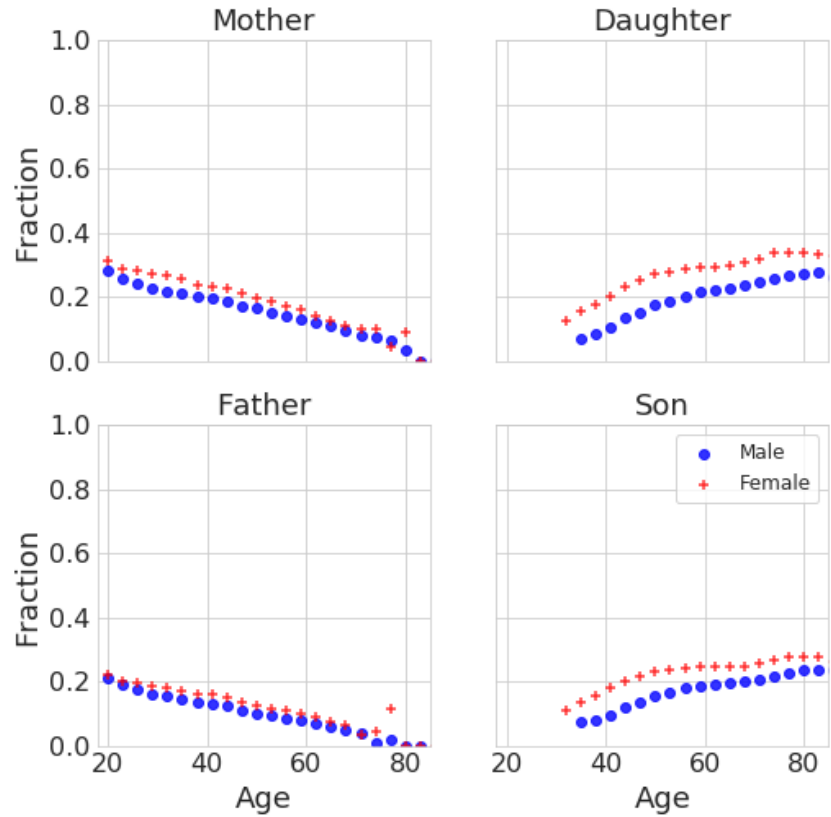

Fig. 2. Ratios of relationship types filtered using the last-names technique from the frequency-age-sex method.

# Statistical Analysis

Table S1[1]. Life-course dependent differences in distribution, mean, and median of kin and non-kin, for every category.

| Variable | Alter type | Ego Sex | Kolgomorov Smirnov | p-val | Kin(mean) | Non-Kin(mean) | Kin(median) | Non-Kin(median) | t-test | Wilcoxon-Mann-Whitney |
|---|---|---|---|---|---|---|---|---|---|---|
| Frequency | Mother | female | 0.37 | 0 | 275.4 | 76.1 | 102 | 14 | 0 | 0 |
| | | male | 0.35 | 0 | 199.5 | 80.8 | 81 | 14 | 0 | 0 |
| | Father | female | 0.34 | 0 | 190.9 | 80.2 | 72 | 12 | 0 | 0 |
| | | male | 0.32 | 0 | 183.3 | 73.2 | 71 | 14 | 0 | 0 |
| | Daughter | female | 0.37 | 0 | 321.5 | 95.6 | 138 | 20 | 0 | 0 |
| | | male | 0.33 | 0 | 226.8 | 106.6 | 98 | 19 | 0 | 0 |
| | Son | female | 0.33 | 0 | 229.7 | 101.8 | 103 | 18 | 0 | 0 |
| | | male | 0.31 | 0 | 217.2 | 96.1 | 96 | 22 | 0 | 0 |
| Fraction of time | Mother | female | 0.38 | 0 | 0.13 | 0.05 | 0.07 | 0.01 | 0 | 0 |
| | | male | 0.35 | 0 | 0.11 | 0.05 | 0.05 | 0.01 | 0 | 0 |
| | Father | female | 0.33 | 0 | 0.09 | 0.04 | 0.04 | 0.01 | 0 | 0 |
| | | male | 0.33 | 0 | 0.09 | 0.04 | 0.04 | 0.01 | 0 | 0 |
| | Daughter | female | 0.38 | 0 | 0.14 | 0.05 | 0.08 | 0.01 | 0 | 0 |
| | | male | 0.33 | 0 | 0.10 | 0.05 | 0.05 | 0.01 | 0 | 0 |
| | Son | female | 0.33 | 0 | 0.10 | 0.05 | 0.05 | 0.01 | 0 | 0 |
| | | male | 0.32 | 0 | 0.09 | 0.05 | 0.05 | 0.01 | 0 | 0 |
| Out-Call Fraction | Mother | female | 0.15 | 0 | 0.49 | 0.46 | 0.49 | 0.43 | 0 | 0 |
| | | male | 0.14 | 0 | 0.49 | 0.48 | 0.50 | 0.48 | 7.92E-54 | 3.18E-88 |
| | Father | female | 0.15 | 0 | 0.44 | 0.45 | 0.42 | 0.40 | 4.06E-06 | 4.37E-37 |

---

[1] For this comparison to be valid, every non-kin group was downsampled considering the size of every Age-Sex-Relationship cohort. For example, If 23 year old mothers are 5230 and non-mothers 21.000, Non mothers were randomly sampled to 5230. This procedure was done, because when age increases, kin detection decreases and non-kin group increases by default and this produces a non-balanced sample.

| | | | | | | | | | | |
|---|---|---|---|---|---|---|---|---|---|---|
| | | male | 0.12 | 0 | 0.45 | 0.46 | 0.43 | 0.45 | 1.46E-53 | 6.90E-05 |
| | Daughter | female | 0.12 | 0 | 0.50 | 0.52 | 0.50 | 0.52 | 1.75E-66 | 3.62E-90 |
| | | male | 0.12 | 0 | 0.56 | 0.54 | 0.57 | 0.57 | 5.31E-38 | 0.051 |
| | Son | female | 0.11 | 0 | 0.50 | 0.50 | 0.50 | 0.50 | 0.116 | 0.134 |
| | | male | 0.10 | 0 | 0.55 | 0.52 | 0.56 | 0.51 | 3.08E-143 | 1.90E-82 |
| Call length | Mother | female | 0.10 | 0 | 105.7 | 95.3 | 69.0 | 58.1 | 9.25E-264 | 0 |
| | | male | 0.08 | 0 | 94.2 | 88.1 | 61.1 | 54.0 | 1.02E-91 | 0 |
| | Father | female | 0.08 | 0 | 90.5 | 86.6 | 60.5 | 53.5 | 1.03E-31 | 0 |
| | | male | 0.07 | 0 | 86.1 | 80.0 | 57.2 | 51.3 | 6.82E-79 | 0 |
| | Daughter | female | 0.08 | 0 | 106.0 | 101.7 | 70.6 | 63.3 | 2.15E-35 | 0 |
| | | male | 0.06 | 4.06E-260 | 89.2 | 91.8 | 60.8 | 58.3 | 3.75E-12 | 6.81E-104 |
| | Son | female | 0.06 | 0 | 95.3 | 94.3 | 62.6 | 58.7 | 0.0049425 | 3.36E-268 |
| | | male | 0.05 | 8.54E-187 | 85.8 | 85.0 | 58.0 | 55.6 | 0.0289988 | 1.51E-108 |

Table S2. Standard deviation of means across time, and total standard deviation of groups.

| Variable | Alter type | Ego Sex | Standard deviation of means across time | | p-value | Total Standard deviation | |
|---|---|---|---|---|---|---|---|
| | | | Kin | Non-Kin | | Kin | Non-Kin |
| Frequency | Mother | female | 69.15 | 23.21 | 7.44E-13 | 437.5 | 243.9 |
| | | male | 68.40 | 23.67 | 1.25E-06 | 323.4 | 263.4 |
| | Father | female | 53.74 | 21.55 | 7.12E-10 | 328.2 | 287.6 |
| | | male | 54.22 | 22.68 | 3.30E-09 | 319.7 | 227.7 |
| | Daughter | female | 43.98 | 17.49 | 7.86E-09 | 475.4 | 273.3 |
| | | male | 30.97 | 22.82 | 2.41E-02 | 359 | 328.2 |
| | Son | female | 25.89 | 18.11 | 1.06E-02 | 350.1 | 294.1 |
| | | male | 30.65 | 18.97 | 1.07E-03 | 350.7 | 264.7 |
| Fraction of time | Mother | female | 0.034 | 0.018 | 1.05E-05 | 0.17 | 0.11 |
| | | male | 0.031 | 0.018 | 6.93E-05 | 0.15 | 0.11 |
| | Father | female | 0.028 | 0.016 | 7.65E-05 | 0.13 | 0.11 |
| | | male | 0.030 | 0.016 | 4.89E-06 | 0.13 | 0.1 |
| | Daughter | female | 0.033 | 0.016 | 1.60E-06 | 0.17 | 0.11 |
| | | male | 0.034 | 0.018 | 4.38E-05 | 0.14 | 0.12 |
| | Son | female | 0.021 | 0.013 | 6.27E-04 | 0.13 | 0.11 |
| | | male | 0.027 | 0.015 | 5.98E-05 | 0.12 | 0.1 |
| Out-Call Fraction | Mother | female | 0.038 | 0.015 | 1.75E-04 | 0.3 | 0.37 |
| | | male | 0.047 | 0.016 | 2.63E-09 | 0.31 | 0.37 |
| | Father | female | 0.054 | 0.022 | 1.31E-09 | 0.3 | 0.38 |
| | | male | 0.058 | 0.021 | 2.23E-11 | 0.3 | 0.37 |
| | Daughter | female | 0.027 | 0.008 | 1.84E-13 | 0.29 | 0.36 |
| | | male | 0.043 | 0.007 | 1.11E-16 | 0.29 | 0.36 |

|  | Son | female | 0.036 | 0.010 | 1.09E-13 | 0.3 | 0.36 |
|  |  | male | 0.052 | 0.010 | 1.11E-16 | 0.29 | 0.35 |
| Call length | Mother | female | 28.22 | 16.74 | 2.45E-10 | 118.4 | 130.3 |
|  |  | male | 21.24 | 8.82 | 1.75E-04 | 106.6 | 123.5 |
|  | Father | female | 29.05 | 12.71 | 1.59E-08 | 99.7 | 121 |
|  |  | male | 23.68 | 11.33 | 3.18E-07 | 95.7 | 107.8 |
|  | Daughter | female | 19.40 | 12.19 | 1.45E-03 | 114.1 | 133 |
|  |  | male | 14.43 | 8.43 | 3.09E-04 | 94.3 | 118.4 |
|  | Son | female | 15.19 | 8.39 | 8.38E-05 | 104.8 | 126.5 |
|  |  | male | 13.79 | 7.33 | 3.29E-05 | 92.6 | 105.6 |

## Data Anonymization and Metadata

The data is aggregated anonymized Call Detail Records (CDRs) from a Chilean Mobile Call Company accounting for ∼40% of the market share for the respective period. CDRs are generated automatically by the telephone company every time that a call is made or received by a person inside the network. This data is collected automatically for billing purposes.

Each record stores, among other things, the origin and destination number and antenna, a time stamp (day, hour, minute, second) and the duration of the call in seconds. The data used in this study was anonymized using several techniques (see the supplementary material for details). First, we used hash functions to convert phone numbers into a different string while still being able to connect all the phone calls that were made by that number. We also excluded attributes from the CDR that are not required or cannot be used for ethical reasons (antennas, exact time of phone calls). The information extracted from CDRs includes phone of origin (Origin Phone), destination phone (Destination Phone) and duration (Duration).

This information was aggregated for every pair of phone numbers that appeared at least once in the CDRs. The final database used was an aggregation of all of the phone calls made during 2015, for which at least one phone was from Movistar as seen in table 1.

| Phone_A | Phone_B | OutCalls | InCalls | Sec |
|---|---|---|---|---|
| 71e61e625c967f98da69 | bbb818a312f0fdb0771d | 3 | 1 | 512 |
| da1f483278cf73d22aa5 | 562a74c3d213871edf6b | 1 | 1 | 333 |
| 71e61e625c967f98da69 | 562a74c3d213871edf6b | 1 | 0 | 957 |

Table 1 Aggregated CDRs

Besides the CDR data, we have an anonymized version of Movistar's Clients Registry with metadata about the owners of the phone lines. This metadata is collected by the company and includes, among other things, date of birth, gender, names, last names and also information about the type of contract (individual or family contract).

For our project the data was restricted to attributes: age (on January 1st2015), gender, first and second last name, and type of contract. For data protection we replaced the date of birth by age on January 1st , 2015. For family contracts we used the detailed records of phone owners given by the company and we only left the oldest phone number with the metadata. The rest of the phones of the plan were left as null values .

Table 2 is an example of the client database with anonymized cell phone numbers (Phone), paternal lastname (LNp), maternal last name (LNm), anonymized owner ID (to identify family and individual plans).

In this example, both phones 71e61e625c967f98da69 and da1f483278cf73d22aa5 have the same owner adfr54kjhy5687lootek. The metadata was always associated to phone that had the oldest contract. In this case 71e61e625c967f98da69.

| Phone | $LN_p$ | $LN_m$ | Sex | Age | OwnerID |
|---|---|---|---|---|---|
| 71e61e625c967f98da69 | X1 | X10 | Male | 42 | adfr54kjhy5687lootek |
| da1f483278cf73d22aa5 | NULL | NULL | NULL | NULL | adfr54kjhy5687lootek |
| 562a74c3d213871edf6b | X24 | X5 | Female | 20 | oujh65dfhk87rfjih677 |
| bbb818a312f0fdb0771d | X24 | X8 | Male | 47 | gp984asw12rcyy998rjh |

Table 2 Client's Registry Database

The data was collected for the 12 consecutive months of the year 2015, totaling 3,994,595,128 calls. Using all this data we created a Mobile Call Graph (Gmc). This directed graph Gmc = (Vmc, Emc) there is a node A ∈ Vmc for every phone number in the dataset, and there is an edge (A, B) ∈ Emc if a, b ∈ Vmc, there is at least one phone call from phone A to phone B. Each node has labels or attributes that correspond to the metadata of the phone. If the phone is not the main one of a Movistar plan or the node does not belong to the Movistar network, all the labels associated to it are null values. The edges are labelled with the number of Out-calls, In-Calls and TotalSec. The final graph created consisted in 8,907,140 vertices and 112,744,511 edges.

One of the aspects that we considered together with Telefonica in the definition of the data to be delivered is that they comply with the handling of personal data required by the GDPR of the European Union. The process described above effectively meets the requirements since the data cannot be attributed to a particular subject without additional information.

**Note on the metadata**

The metadata contain also information about the type of telephone contract (individual or family contract). We only used nodes where metadata was available for both egos and at least one of the alters. For family contracts we used the detailed records of phone owners given by the company. If an individual moved from a private phone contract to a family contract, it was assumed that the person kept using the same number instead of switching the phone with another member of the family. Thus, for each family contract we only left one node, the node where ownership of the phone can be traced to the person signing the family contract. The rest of the phone numbers included in the family contract were not included in the analysis left as grey nodes f(no metadata).